\documentstyle[12pt]{article}
  \advance\oddsidemargin  by -1in
  \advance\evensidemargin by -1in
\def\lsim{\mathrel{\rlap{\lower4pt\hbox{\hskip1pt$\sim$}}
    \raise1pt\hbox{$<$}}}         
\def\gsim{\mathrel{\rlap{\lower4pt\hbox{\hskip1pt$\sim$}}
    \raise1pt\hbox{$>$}}}         

\def\be{\begin{equation}}
\def\ee{\end{equation}}
\def\bq{\begin{eqnarray}}
\def\eq{\end{eqnarray}}

\begin{document}
\pagestyle{empty}
\hfill{\large DFTT 61/93}

\hfill{\large October 1993}

\vspace{2.0cm}

\begin{center}

{\large \bf SHADOWING IN DEUTERIUM \\
AND THE SMALL--$x$ LIMIT\\
OF $F_2^n/F_2^p$ AND $F_2^p - F_2^n$ \\}
\vspace{1.0cm}
{\large V.~Barone$^{a}$, M.~Genovese$^{a}$ ,
N.N.~Nikolaev$^{b,c}$,\\
E.~Predazzi$^{a}$ and B.G.~Zakharov$^{b}$ \vspace{1.0cm} \\}
{\it
$^{a}$Dipartimento di
Fisica Teorica, Universit\`a di Torino\\
and INFN, Sezione di
Torino,    10125 Torino, Italy\medskip\\
$^{b}$L. D. Landau Institute for Theoretical Physics, \\
GSP-1,
117940,
                Moscow V-334, Russia \medskip \\
$^{c}$IKP (Theorie), KFA J{\"u}lich,
5170 J{\"u}lich, Germany}
                \vspace{1.0cm}\\

{\large \bf Abstract \bigskip\\ }

\end{center}

We discuss the updated NMC determination of
$F_2^n/F_2^p$ and $F_2^p - F_2^n$.
Shadowing effects
in deuterium
make the structure functions  determined by the NMC
sensibly different from the true ones in the low--$x$ region.
We show that the departure
of $F_2^n/F_2^p$ and
$F_2^p-F_2^n$ from the Regge expectations at small $x$
observed by the NMC
likely disappears if one takes into account
the shadowing corrections.

\pagebreak
\pagestyle{plain}

The New Muon Collaboration has recently presented \cite{NMC3}
a re--evaluation of the ratio $F_{n/p} \equiv F_2^n/F_2^p$ and of the
difference $F_{p-n} \equiv F_2^p - F_2^n$,
which leads to a new value of the
Gottfried sum $S_G = \int {\rm d} x \, (F_2^p - F_2^n) /x$,
slightly larger
than the result previously published \cite{NMC1}.

The novel finding we shall
focus on is the small--$x$
behavior of $F_{n/p}$, which
appears now to be
definitely
smaller than unity at the lowest $x$--bin.
The occurrence of this fact was anticipated or implicitly
expected \cite{BGNPZ,Z,BK,NZ}
as a possible evidence
of shadowing in deuterium. The purpose of the present paper
is to endorse quantitatively such a prediction, showing
how the NMC results should be corrected in order to
account for nuclear effects in deuterium.
Our results will be obtained in the framework of the
shadowing model of Ref.~\cite{NZ,BGNPZ}.

Since the NMC results are the output of a rather complicated
procedure, which starts from the direct measurements
of the structure functions and involves fits of the data
from other experiments and some simplifying assumptions, it is
useful to summarize the method used by the NMC
\cite{NMC3,NMC1,NMC2} to derive
$F_{n/p}$ and $F_{p-n}$.

The cross sections for deep inelastic scattering off deuterium ($D$)
and hydrogen are simultaneously measured
and the ratio $F_{n/p}$ is then determined neglecting
shadowing in deuterium
\be
F_{n/p}^{NMC} = \frac{2 F_2^D}{F_2^p} - 1\,.
\label{eq1}
\ee
Here $F_2^D$ is the deuterium structure function per nucleon and
the superscript $NMC$ will denote hereafter the quantities
obtained under the assumption of no shadowing in deuterium
 (which is the strongest assumption made by the NMC in its data
analysis).

In order to apply the radiative corrections to $F_{n/p}$, the
knowledge of both $F_2^p$ and $F_2^D$ is required. The latter is obtained
from a fit to the available data (including those from other
experiments), while $F_2^p$ is derived from the $F_2^D$ fit
and the measured ratio $F_{n/p}$, inverting eq.~(\ref{eq1}).
The procedure is iterated for few steps until the change
in $F_{n/p}$ becomes reasonably negligible.

Once $F_{n/p}$ is determined, along the lines illustrated above, the
non--singlet structure function $F_{p-n}$ is obtained as
\be
F_{p-n}^{NMC} = \frac{1 - F_{n/p}^{NMC}}{1 + F_{n/p}^{NMC}}
\, 2 F_2^D\,,
\label{eq2}
\ee
where $F_2^D$ is again given by a global parametrization.

The difference between the most recent NMC determination of
$F_{n/p}$ and the previous one \cite{NMC2} resides in
the $F_2^D$ fit which has been updated and gives now larger values
of $F_2^D$ at low $x$. The latest NMC results for
$F_{n/p}$ and $F_{p-n}$ at $Q^2 =
4 \, GeV^2/c^2$ are displayed in Figs.~1 and 2, respectively,
and show that, at small $x$, $F_{n/p}$ tends to a value
definitely smaller than unity.

In the limit $x \rightarrow 0$ the structure functions are expected
to exhibit the Regge behavior \cite{books}.
In particular, the sea parton densities
are dominated by the pomeron pole, which is isospin and flavor independent.
Thus

\be
F_{n/p} \rightarrow 1\,,\;\;
F_{p-n} \rightarrow 0
\;\;\;\;
{\rm as} \,\, x \rightarrow 0\,.
\label{eq3}
\ee
Independently of the Regge behavior, the same limiting values
would be obtained  by assuming
both the flavor $SU(2)_f$ and the isospin $SU(2)_I$
symmetries to be valid.
We recall that
the experimental result for the Gottfried sum $S_G$ \cite{NMC3,NMC1},
much smaller than the naive expectation $1/3$ seems to indicate
that, at least at intermediate $x$, either $SU(2)_f$ and/or
$SU(2)_I$ are broken (see discussion and references in \cite{F}).

Notice that the expectations above are valid
for the {\it true}
structure functions.

If the measured
ratio $F_{n/p}$ is different from unity  at some small $x$ value
(as it was found by the NMC), one can conceive two
explanations:
{\it i)} the pomeron--dominated region has not
been reached and either $SU(2)_f$ and/or $SU(2)_I$ are violated
at small $x$ (for instance, by flavor-- and/or isospin--dependent
terms of the form
$\sim x^{\alpha}$, with $-1 <\alpha <0$)
 ; {\it ii)} due to shadowing in
deuterium, the measured structure functions are
{\it not} the true structure functions,
and in this case the Regge behavior might still hold.

While it is not excluded a priori that the first possibility
can contribute to the effect, the second one
is to be investigated in order
to evaluate how much the measured quantities should be corrected.
Shadowing is undoubtedly at work in deuterium and, as a matter of
fact, we shall see that it leads to sensible corrections
to the experimental results, which point towards the
asymptotic values (\ref{eq3}).
We also mention that pionic corrections have been shown \cite{Z2} to
be unable to explain the departure of the experimental data
on $F_{n/p}$ and $F_{p-n}$ from
the conventional expectations at small $x$ (pions affect the
$n/p$ ratio only in the valence region).

At this point, let us introduce the shadowing ratio
\be
R_s(x) = \frac{2 \, F_2^D(x)}{F_2^n(x) + F_2^p(x)}
\label{eq4}
\ee
which was calculated in \cite{NZ,BGNPZ} and is about $3 \%$ different
from unity at $x = 7 \cdot 10^{-3}$ and $Q^2 =  4 \, GeV^2/c^2$.
In terms of $R_s$, the true ({\it i.e.}
corrected) ratio $F_{n/p}^{true}$ becomes
\be
F_{n/p}^{true} = \frac{2 \, F_2^D}{R_s F_2^p} - 1
= \frac{1}{R_s} ( F_{n/p}^{NMC} - R_s + 1)\,,
\label{eq5}
\ee
and is obviously larger than $F_{n/p}^{NMC}$.

In an analogous way, one finds for the true difference
$F_{n-p}^{true}$
\be
F_{p-n}^{true} = \frac{1 - F_{n/p}^{true}}{1 + F_{n/p}^{true}}
\, \frac{2 \, F_2^D}{R_s} =
\frac{2 R_s - (1 + F_{n/p}^{NMC})}{R_s(1 - F_{n/p}^{NMC})} \,
F_{p-n}^{NMC}\,.
\label{eq6}
\ee
which is smaller than $F_{p-n}^{NMC}$.

In order to give a quantitative estimate of the difference between
NMC and true quantities, we resort to the model calculations
of Ref.~\cite{BGNPZ} (for details we refer the reader
to that paper). It is possible
to cast the results for $R_s(x,Q^2)$ obtained in \cite{BGNPZ}
in the following form, valid in the range $1\, GeV^2/c^2  \le Q^2
\le 10 \, GeV^2/c^2\,,\; x \gsim  10^{-3}$
\be
1 - R_s(x,Q^2) = A(Q^2) \, (1 - x/x_0)^{B(Q^2)}\,,
\label{eq7}
\ee
where $x_0 = 0.2$ and $A(Q^2) = 0.043 - 0.0049 \, \ln (Q^2/Q_0^2)\,,
Q_0^2= 1\, GeV^2/c^2; \;
B(Q^2) = 9.83 - 0.13 \,Q^2$.

The ratio $R_s(x,Q^2)$ is shown in Fig.~3 for three different values
of $Q^2$. At $Q^2 = 4 \, GeV^2/c^2$, which is the value of the NMC
results, the shadowing is a $3 \%$ effect at $x = 5 \cdot 10^{-3}$
and rapidly falls down, vanishing at $x \simeq 0.1$. However its effect
on $F_{n/p}$ and $F_{p-n}$ is not negligible, even within the
present experimental errors.

Inserting the ratio $R_s(x, Q^2= 4 \, GeV^2/c^2)$ in
eqs.~(\ref{eq5},\ref{eq6}) and taking as input the NMC data
we get for $F_{n/p}^{true}$ and $F_{p-n}^{true}$ the results
shown in Figs.~1 and 2. As one can see, the corrections due
to shadowing are relatively important and raise (lower) the
value of $F_{n/p}$ ($F_{p-n}$). The corrected $n/p$ ratio
tends to unity at the smallest $x$ value, whereas the
$p-n$ difference tends to zero. No apparent departure from the
Regge behavior is thus detected after the corrections.
To account for the theoretical uncertainties
in the calculation of shadowing we varied $1-R_s$ by $ \pm 30 \%$.
The effect of this variation
is given by the two dotted curves in Figs.~1 and 2
and does not fill the discrepancy between the measured and the
corrected quantities.

It has been already pointed out \cite{BGNPZ,Z,BK,BGNPZ2} that
the shadowing corrections sensibly affect the {\it true}
value of the Gottfried
sum, which turns out to be lower than the one quoted by the NMC
\cite{NMC3,NMC1}. In our model we find that $S_G^{NMC}$
should be decreased,
in the measured region
$x > x_{min} = 0.004$, by
$\delta S_G(x_{min},1) \simeq  0.03$.

In conclusion, we have evaluated the corrections to $F_{n/p}$
and $F_{p-n}$ due to the presence of shadowing in deuterium.
We stress that it is important to account for these corrections
in the global parametrizations
of the deep inelastic data which make use of the $n/p$ ratio.
We have also seen that, allowing for shadowing,
there is at present no evidence for a violation of the
conventional Regge expectations.

\vspace{2cm}
{\large \bf Acknowledgment.} It is a pleasure to thank
Stefano Forte
for interesting discussions.

\pagebreak

\pagebreak

\begin{center}

{\large \bf Figure captions}

\end{center}

\vspace{1cm}

\begin{itemize}

\item[Fig.~1]
The structure function ratio $F_2^n/F_2^p$ at small $x$ and $Q^2
= 4 \, GeV^2/c^2$. The squares are the NMC results \cite{NMC3}, the
circles are the central values of the data
data corrected for shadowing. The dotted lines
represent a  $\pm 30 \%$ variation of the amount of shadowing in
deuterium (see text).

\item[Fig.~2]
Same as Fig.~1 but for the structure function difference $F_2^p- F_2^n$.

\item[Fig.~3]
The quantity $1- R_s$ (see text) calculated in the model of
Ref.~\cite{NZ,BGNPZ} for three different $Q^2$ values (dashed line:
 $1 \, GeV^2/c^2$; solid line: $4 \, GeV^2/c^2$; dotted line: $10 \,
GeV^2/c^2$).

\end{itemize}

\end{document}